\documentclass[preprint,showpacs]{revtex4}
\usepackage{epsfig}
\begin{document}

\title{Noise Induced Complexity: From Subthreshold Oscillations to
  Spiking in Coupled Excitable Systems}

\author{M. Zaks\email{Zaks@physik.hu-berlin.de}, X. Sailer, and L.
  Schimansky-Geier}
\affiliation{ Institut f\"ur Physik, Humboldt-Universit\"at zu Berlin,
  Newtonstr.~15, D-12489 Berlin, Germany }
\author{A. Neiman} \affiliation{Department of Physics and Astronomy,
  Ohio University, Athens, OH 45701, USA}

\date{\today}
\begin{abstract}  
  We study stochastic dynamics of an ensemble of $N$ globally coupled
  excitable elements. Each element is modeled by a FitzHugh-Nagumo
  oscillator and is disturbed by independent Gaussian noise. 
  In simulations of the Langevin dynamics we characterize the collective 
  behavior of the ensemble in terms of its mean field and show that with the 
  increase of noise the mean field displays a transition from a steady equilibrium to 
  global oscillations and then, for sufficiently large noise, back to another 
  equilibrium.  Diverse regimes of collective dynamics ranging
  from periodic subthreshold oscillations to large-amplitude
  oscillations and chaos are observed in the course of this transition.
  In order to understand details and mechanisms of noise-induced
  dynamics we consider a thermodynamic limit $N\to\infty$ of the ensemble,
  and derive the cumulant expansion describing temporal evolution of the 
  mean field fluctuations. In the Gaussian approximation this allows us to
  perform the bifurcation analysis; its results are in good agreement 
  with dynamical scenarios observed in the stochastic simulations of 
  large ensembles. 
\end{abstract}

\pacs{05.40-a, 05.45.Xt}
\maketitle
%
%
{\bf Recent studies have shown that noise may greatly enrich dynamics
  of nonlinear systems. This contradicts an intuitive conception on a
  simple blurring effect of noise added to a dynamical system. Thus,
  the quantitative measures of noise become additional control
  parameters of the system, extending its parameter space. For
  instance, variation of noise intensity may lead to qualitative
  transformations in the spatio-temporal dynamics. Thus noise changes
  the complexity of a system.  As an example of such behavior we
  present an ensemble of globally coupled excitable elements. We let
  Gaussian white noise act additively and independently on each
  element of the ensemble.  Simulations of the coupled Langevin
  equations as well as the bifurcation analysis of the equations for
  the lowest five cumulants show a rather complex sequence of
  qualitative changes when the intensity of the additive noise is
  varied.}

\section{Introduction}
Noise-enhanced order in nonlinear dynamics far from equilibrium has
become a rapidly growing area of modern statistical physics. Stochastic
resonance \cite{gammaitoni98}, directed fluxes in stochastic ratchets
\cite{linke02}, noise-induced phase transitions \cite{garcia99}
are intensely discussed phenomena in nonlinear stochastic
systems. Common to all of them is that an increase of the noise level  
results in a more ordered spatio--temporal response.
This growth of order can be expressed and measured using various quantities.
In particular, application of information theory allowed to quantify directly 
the noise-enhanced ordering and information transfer in bistable and excitable 
systems \cite{entro-prl,hanggi-igor-pre,igor-pre}.

A broad class of dynamical systems with diverse applications 
is characterized by excitable dynamics. Chemical reactions
\cite{sakurai02}, lasers \cite{wuensche01},
models of blood clotting \cite{lobanova04}, cardiac tissues
\cite{panfilov00} and nerve cells \cite{koch99}
belong to its most prevalent realizations.  An excitable system
possesses a stable equilibrium state (or rest state) from which it can
temporarily depart if disturbed by a large enough stimulus: the system
responds with a spike, a large excursion in its phase space, returning
back to the rest state through a refractory period. Fundamental
phenomena in excitable systems, such as pulse propagation, spiral
waves, spatial and temporal chaos and synchronization are well studied
\cite{murray03,mikhailov94,chay85,Hu01mitZhou}.

Analysis of the influence of noise on excitable dynamics has recently attracted
great attention \cite{kadar98,jung98,longtin98,bahar02,hu01,zhou03,kanamaru01,schmid03,schmid04,lindner04,acebron04,hanggi-neuron-prl}
as a necessary step towards realistic description of natural systems.
The inclusion of fluctuations into models of excitability is desirable
due to their omnipresence in processes of signal transmission and
detection \cite{tuckwell88} which constitute often an interplay of
signals and noise.  Hence, the strength of variability as a measure of
the present noise enters the excitable models as a {\em new}
parameter \cite{lindner04}.  That is why excitable dynamical systems 
{\em per~se}  are in nonequilibrium and the acting noise is unbalanced with
respect to dissipative forces. Hence, a variation of noise often may
induce qualitative changes in the performance and functioning of
excitable systems which can be characterized in terms of the dynamics
of moments and their bifurcations
\cite{malakhov74,rodriguez96,hasegawa03,tanabe01,kawai04,zaks03}.

A variety of models was proposed to describe different regimes of
excitable systems, including dynamics of excitable, spiking, and
bursting neurons \cite{izhikevich00}. Here we concentrate  on
one of the most popular systems, the FitzHugh-Nagumo model
which was derived from the Hodgkin-Huxley
model of the giant squid axon \cite{mikhailov94}. Despite its relative
simplicity it displays a variety of biologically relevant dynamical
regimes if taken as a single element or embedded into a network.  It
has been shown that the influence of noise on excitable systems such
as the FitzHugh-Nagumo model can lead to a coherent temporal response via
the phenomena of stochastic \cite{gammaitoni98} and
coherence \cite{lindner04,pikovsky97,shinohara02} resonance.

In this paper we study noise-induced dynamics in a network of globally
coupled FitzHugh-Nagumo elements, each with its own noise source. 
Most studies on noise-induced transitions considered so-called multiplicative 
or parametric noise or noise being colored \cite{garcia99}. In contrast, in our
case excitable elements are driven by {\em additive} Gaussian 
{\em white} noise.  Being statistically independent in different 
elements, noise sources possess the same intensity, $T$, which we use
as a control parameter. We concentrate on the parameter region where a
single uncoupled element possesses a stable equilibrium and is excitable.
We simulate the stochastic equations and show that variation of 
$T$ results in a rich variety of regimes of the collective response
the ensemble, expressed in terms of the mean field (Section \ref{chapLangevin}). 
In order to delineate dynamical mechanisms behind
these transitions we use cumulant approach in Gaussian approximation,
thereby passing from a description in terms of coupled Langevin
equations to a deterministic system of the fifth order for the
cumulants of the ensemble distribution. This
system is further analyzed by methods of the bifurcation theory 
(Section \ref{chapCumulants}). In particular, separation of timescales 
allows  to identify the transition from irregular minor oscillations
to regime of intermittent spikes of large amplitude as the canard
explosion of a chaotic attractor.  We demonstrate that the spiking regime,
whether regular or irregular, is possible: (a) only in the restricted
interval of noise intensities, and (b) for moderate values of the coupling
strength: a too strong coupling holds the ensemble together and keeps it
close to the equilibrium; a too strong noise maintains the uniform distribution
with non-oscillating mean values.

\section{Langevin approach: numerical simulation}
\label{chapLangevin}
As a prototype of an ensemble of excitable units we investigate the
set of $N$ noise-driven FitzHugh-Nagumo systems.  The individual
systems are coupled through the mean field and obey the equations:
\begin{eqnarray}
\epsilon\dot{x}_i&=&x_i-\frac{x_i^3}{3}-y_i
       +\gamma\,(\langle x\rangle -x_i) ,\nonumber \\
\dot{y}_i&=&x_i+a+\sqrt{2T}\,\xi_i(t), \hspace{1.cm}  i=1,\ldots,N.
\label{fhn}
\end{eqnarray}
In the biochemical context the variables $x_i$ denote e.g. the values of the 
membrane potential whereas $y_i$ are responsible
for the inhibitory action. 
Here $\xi_i(t)$ are the independent ($\langle
\xi_i(t)\xi_j(t)\rangle=\delta_{ij}$) white Gaussian noise sources
with vanishing average which individually act on the recovery
variables $y_i(t)$.  This could be interpreted, e.g., as a random
opening of ion channels \cite{falcke04} which stochastically changes the
conductivity of a membrane \cite{zaks05}.

In the absence of noise
($T=0$) the system (\ref{fhn}) possesses a unique equilibrium
state with $x_i=-a$ and $y_i=a^3/3-a$.  The equilibrium is stable for
$a^2>{\rm max}(1,1-\gamma)$ and unstable otherwise. This follows from
the characteristic equation which reads
$$\left(\lambda^2-\frac{1-a^2}{\epsilon}\,\lambda+\frac{1}{\epsilon}
\right)\left(\lambda^2
  -\frac{1-a^2-\gamma}{\epsilon}\,\lambda+\frac{1}{\epsilon}\right)^{N-1}=0$$
and means that positive coupling strength $\gamma$, whatever large it
is, does not influence the stability of equilibrium.

As instantaneous global characteristics of the set of FitzHugh-Nagumo
systems we consider the first moments of the distribution: mean field
components $\langle x\rangle(t)$ and $\langle y \rangle(t)$.

The evolution equations (\ref{fhn}) were solved numerically for the
ensemble of $N=10^5$ oscillators at different values of the noise
intensity $T$.  For our computations, we take $\epsilon=0.01$,
separating thereby the characteristic timescales of the variables.
Further, we choose the value $a=1.05$ under which an individual system
is excitable: if initial perturbation exceeds a certain threshold
value, the system ``fires'' i.e. displays a spike of large amplitude
before finally settling down to the steady state.  Since
$(1-a^2)^2<4\epsilon$, the leading eigenvalues, which characterize the
perturbation decay in the immediate vicinity of the equilibrium, are
complex.

We begin with the moderate value of the coupling strength 
$\gamma=0.1$ and investigate the temporary patterns of the mean
fields for different values of the noise intensity $T$.

As soon as the noisy perturbations are introduced, the individual elements, 
from time to time, are kicked across the excitation threshold and  
exhibit spikes. However, under sufficiently low values of $T$
such irregular spikes are very seldom and do not contribute to the mean fields.
Of course, since the noise is white, the nearly simultaneous ``firing''
of large subsets of elements should occur for arbitrarily
small $T$, but such events appear to be extremely rare, and 
we never observed them within the integration intervals 
of $t\leq 10^3$. 

\begin{figure}[h]
\centerline{\epsfxsize=1.0\textwidth\epsfbox{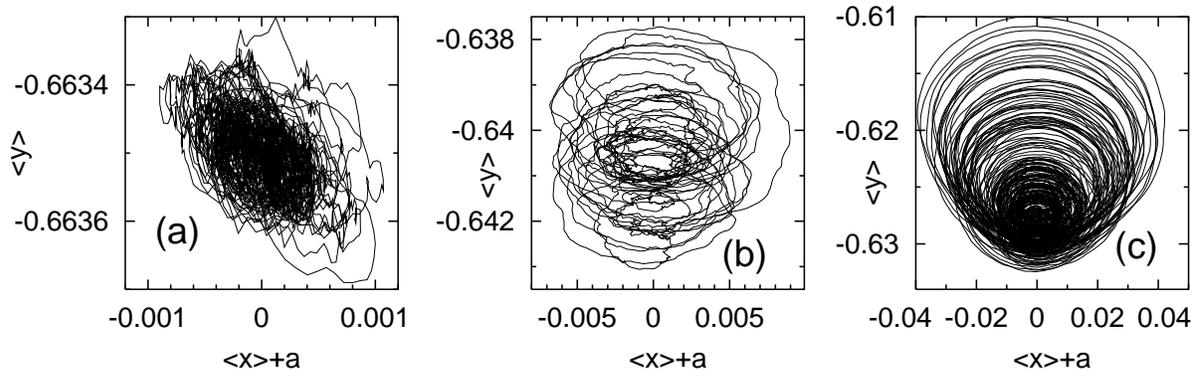}}
\caption{Noise-induced onset of local oscillations around the equilibrium 
    for $a=1.05$, $\epsilon=0.01$, $\gamma=0.1$, $N=10^5$: 
    (a) $T=10^{-4}$,\ \ (b)
    $T=2.4\times10^{-4}$,\ \ (c) $T=2.7\times10^{-4}$. Changes in the
    phase portrait indicate a bifurcation from disordered fluctuations
    to subthreshold oscillations. Note the smallness of the oscillation 
    amplitude.}
\label{triport}
\end{figure}

As seen in Fig.\ref{triport}, at low noise intensities the
mean field exhibits only minor excursions from some average value. For
$\langle x\rangle$ the time average virtually does not decline from
the equilibrium state $-a$ whereas the time average of $\langle
y\rangle$ gets shifted.  The amplitude of these excursions grows with
the increase of $T$. Besides, qualitative changes occur: for very
small $T$ dynamics of the mean field is akin to a non-biased random
walk (Fig.\ref{triport}(a)), at slightly larger values of $T$ the
motion gradually acquires the character of noisy rotation
(Fig.\ref{triport}(b)), and still higher values of $T$ yield phase
portraits which are reminiscent of the spiral chaotic attractor
(Fig.\ref{triport}(c)). In order to exclude the numerical artefacts
and diminish the finite-size effects we performed several control runs
with a much higher ($N=3.2\times 10^7$) number of coupled systems and
obtained in the latter case the even more pronounced pattern of the
multi-band chaotic attractor.

\begin{figure}[h]
\centerline{\epsfxsize=0.75\textwidth\epsfbox{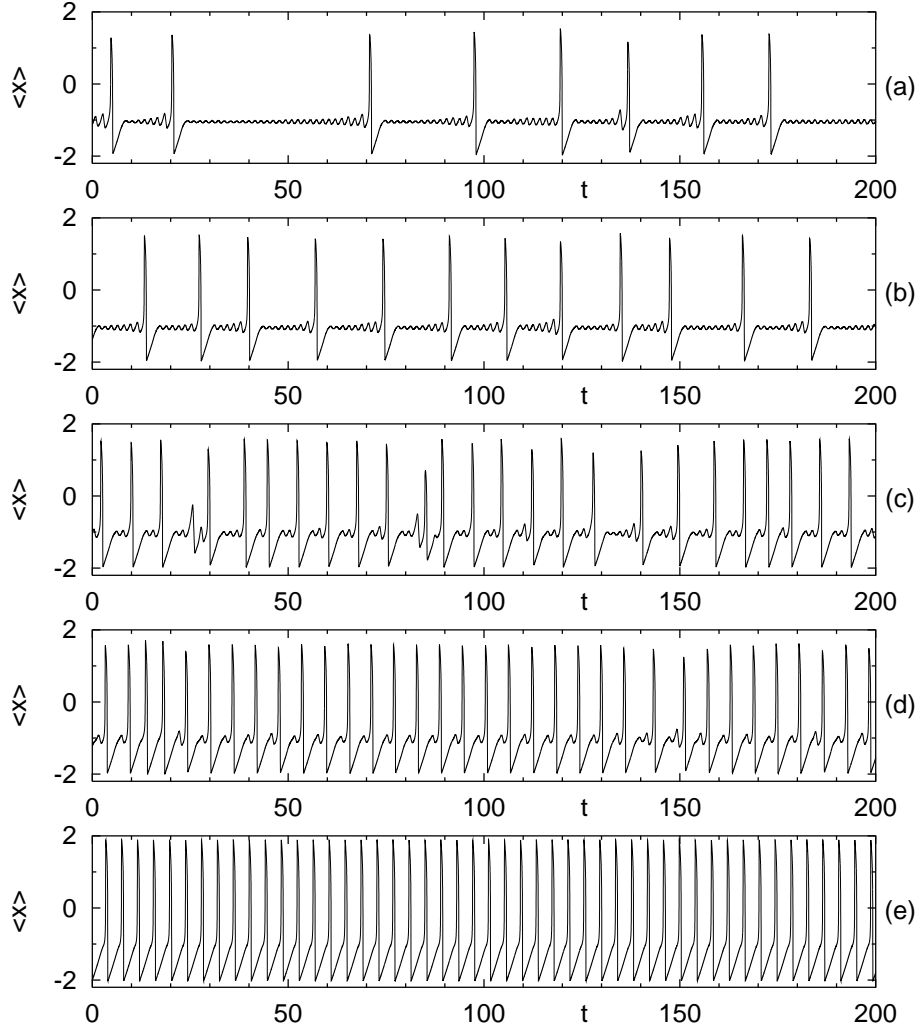}}
\caption{Transition from intermittent spiking of the mean field to the
  regular spiking pattern. The sequence of pictures shows an increase
  of coherence in the response of the ensemble with respect to
  increasing noise. The noise intensity is varied:
  (a)~$T=2.77\protect\times 10^{-4}$,
  (b)~$T=2.8\protect\times10^{-4}$,\ \ (c)~$T=2.9\protect\times
  10^{-4}$, (d)~$T=3.0\protect\times10^{-4}$,\ \ 
  (e)~$T=3.1\protect\times 10^{-4}$, other parameters like in Fig.
  \ref{triport}. }
\label{spiking}
\end{figure}

Starting from $T\approx 2.76\times 10^{-4}$ we observe the qualitative
transition: minor (``subthreshold'') oscillations of the mean field
are interrupted by large outbursts (spikes). This is a consequence of
the collective phenomenon: for a large proportion of the individual
units the firing events occur almost synchronously.  Immediately
beyond the threshold the spiking of the mean field is intermittent:
the length of time intervals between the spikes is strongly varying
(Fig.\ref{spiking}a). With the increase of $T$ the interspike
intervals become shorter and more homogeneous (Fig.\ref{spiking}b).
The number of minor oscillations between two subsequent spikes
steadily decreases (Fig.\ref{spiking}c), proceeding from very large
values (dozens in Fig.\ref{spiking}a) to 2-3 in Fig.\ref{spiking}c,
and 1-2 in Fig.\ref{spiking}d. Finally, such minor oscillations
completely disappear and the mean field exhibits frequent
non-interrupted spiking (Fig.\ref{spiking}e).  Hence variation of
noise creates a coherence resonance of the coupled ensemble.

\begin{figure}[h]
\centerline{\epsfxsize=0.9\textwidth\epsfbox{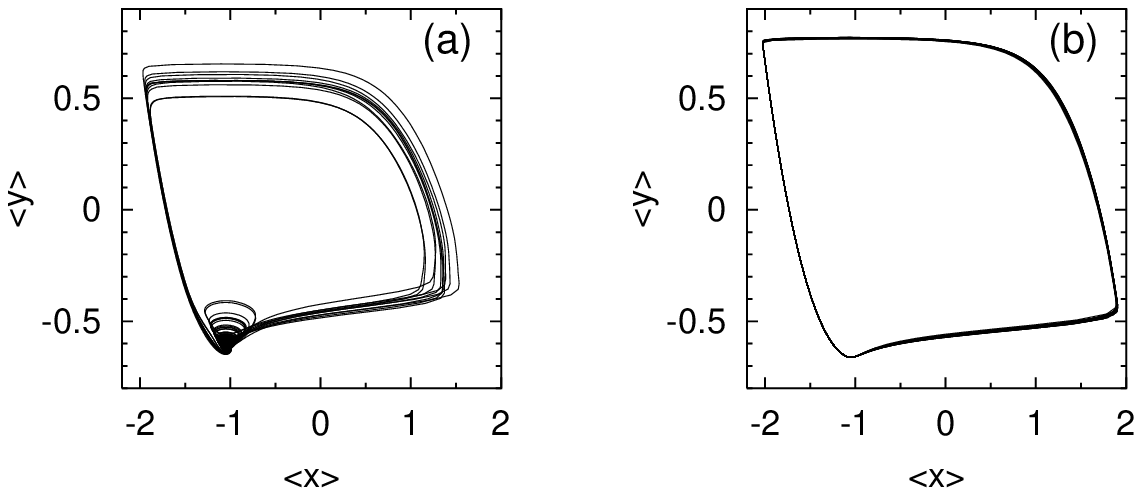}}
\caption{Phase portraits of the mean field: (a)~Intermittent spiking with  
    $T=2.77\protect\times10^{-4}$ , (b)~Regular oscillations with
    $T=3.1\protect\times10^{-4}$, other parameters like in
    Fig. \ref{triport}.}
\label{ph_por_mean}
\end{figure}

The evolution of the phase portrait for the mean field is presented in 
Fig.\ref{ph_por_mean}:
the attractor which corresponds to the state of intermittent spiking
consists of small and large loops (Fig.\ref{ph_por_mean}a);
the attractor of the non-interrupted spiking state is almost indistinguishable
from the large-scale limit cycle (Fig.\ref{ph_por_mean}b).

\begin{figure}[h]
\centerline{\epsfxsize=0.6\textwidth\epsfbox{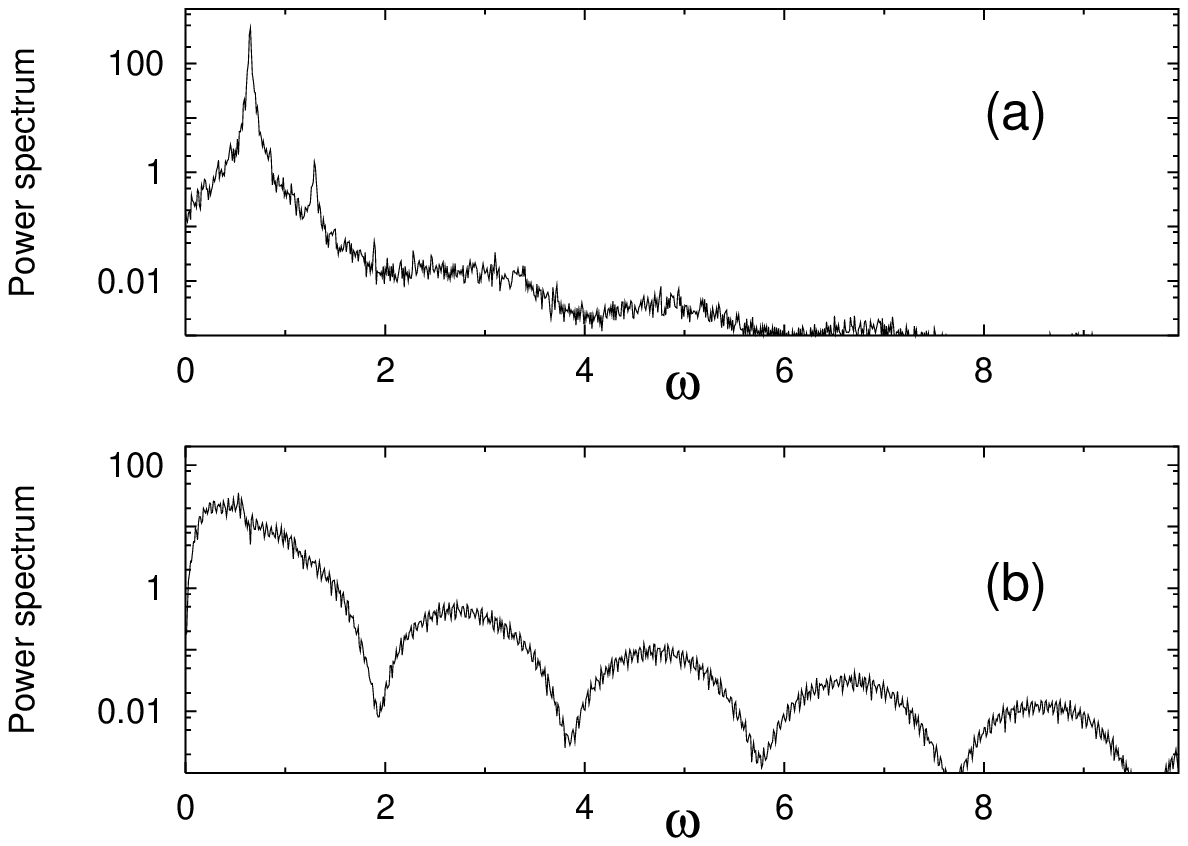}}
\caption{Power spectra of the mean field $\langle x\rangle (t)$: 
(a)~Regime of irregular subthreshold oscillations at
$T=2.73\protect\times10^{-4}$, (b)~Regime of intermittent spiking at
$T=2.76\protect\times10^{-4}$, other parameters like in
Fig.\ref{triport}. Minima in the spectral curve
correspond to the inverse recovery time.}
\label{dir_spectra}
\end{figure}

Noteworthy, transition to the spiking regime is accompanied by drastic
changes in the power spectrum of the mean field. Just below this
transition the spectrum possesses a well-defined peak which sits upon
the noisy background and corresponds to the typical frequency of one
minor rotation around the equilibrium (Fig.\ref{dir_spectra}a). Above
the transition the mean field is characterized by the broadband
spectrum (Fig.\ref{dir_spectra}b); the latter peak can still be
vaguely identified, but there is no sharp contrast to the neighboring
frequency values. A new feature of the power spectrum are the
characteristic deep minima at the value which corresponds to the
inverse ``recovery time'' as well as at its harmonics \cite{prager03}.

In fact, this minima, albeit not so pronounced, can be recognized already
in Fig.\ref{dir_spectra}a. This is explained by the fact that 
the individual units of the ensemble exhibit intermittent spiking even before 
the transition of the mean field. Close before the threshold 
small clusters of such elements start to spike cooperatively,  
making thereby the inverse recovery time visible in the spectrum
already at this stage.

\begin{figure}[h]
\centerline{\epsfxsize=0.4\textwidth\epsfbox{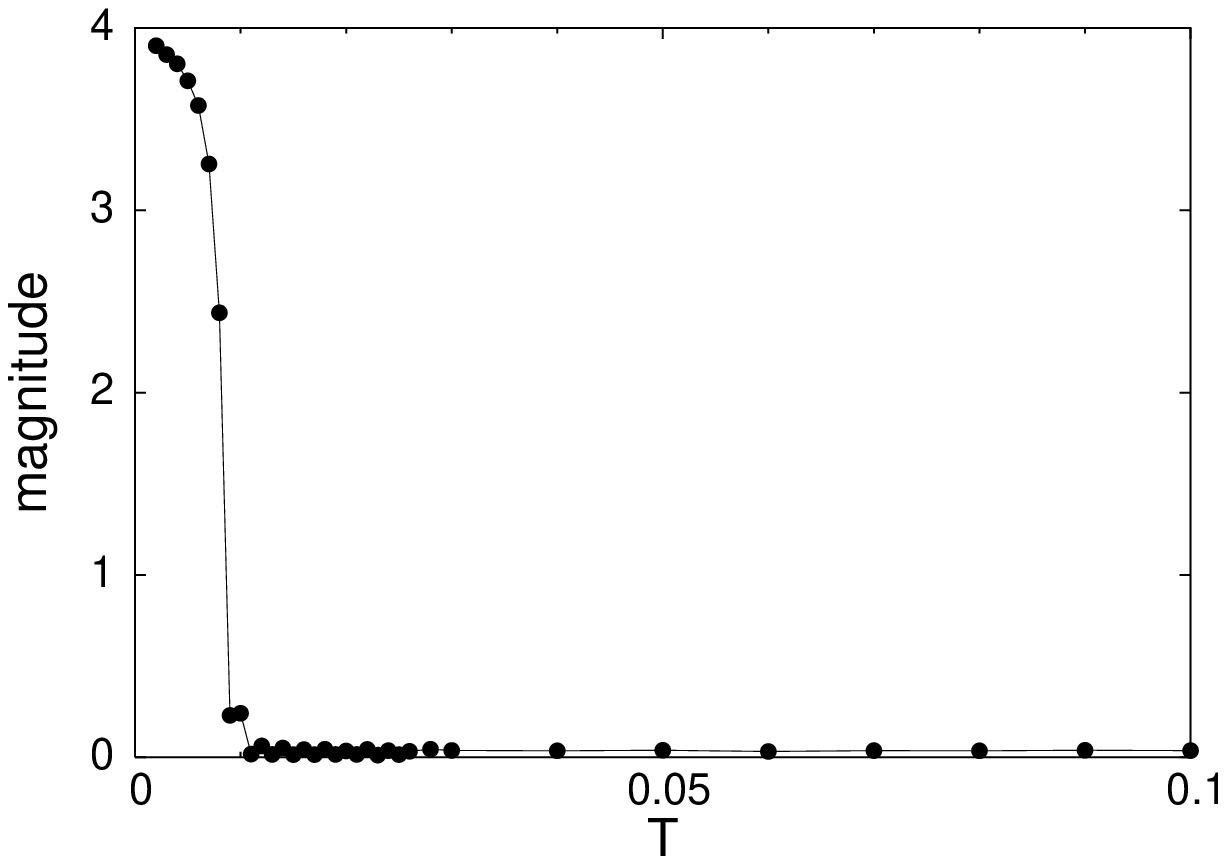}\quad
\epsfxsize=0.4\textwidth\epsfbox{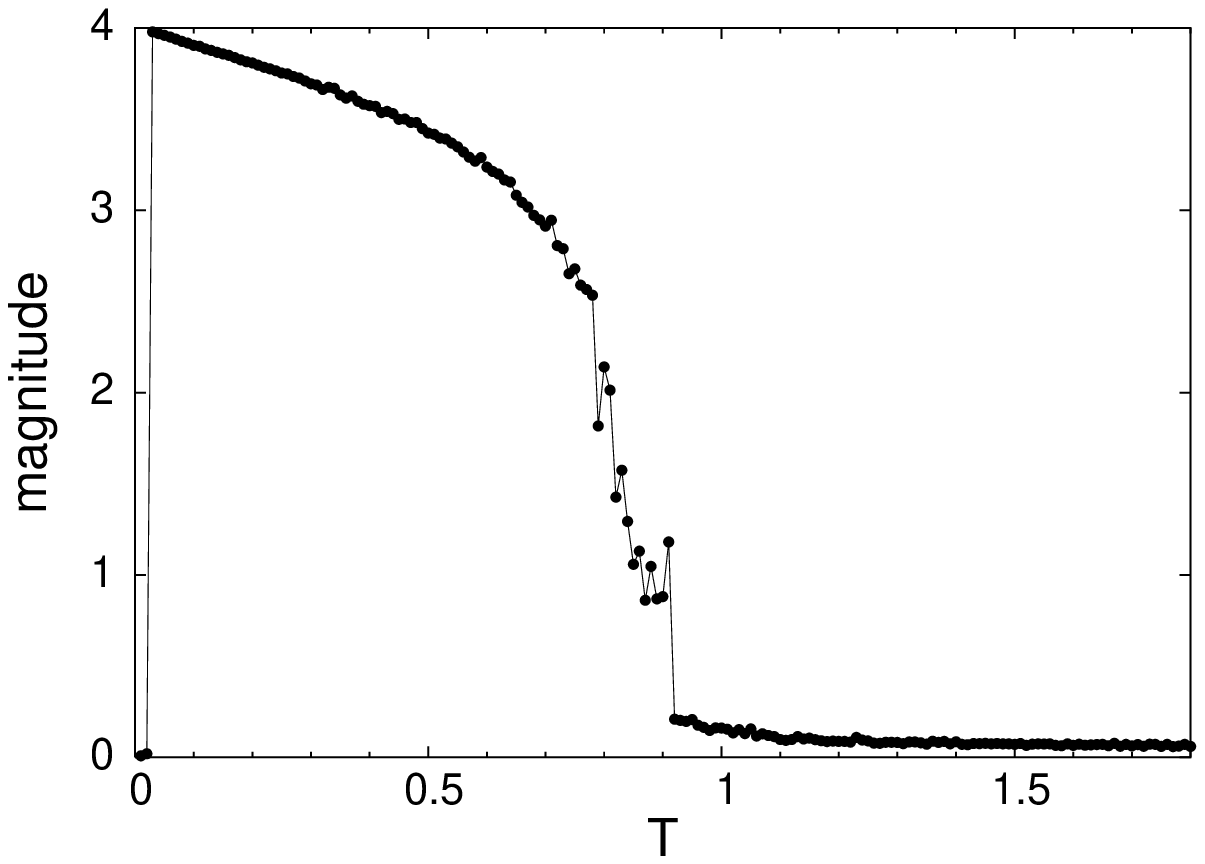}}
\caption{Resonance-like dependence of the magnitude of oscillations 
  on the noise intensity $T$ for different coupling strength. Coupling
  enlarges the range of noise intensities where the resonance occurs.
  Left panel: $\gamma$=0.1; right panel: $\gamma$=1, other parameters
  like in Fig. \ref{triport}.}
\label{dir_diam}
\end{figure}

The range of noise intensities which enable the regime of spiking is
relatively narrow. It is convenient to characterize the oscillatory
states by means of the magnitude: distance between the global extrema
of one of the mean fields, i.e.  $d\equiv \max(\langle x\rangle) -
\min(\langle x\rangle)$.  In Fig.\ref{dir_diam}a we present the
dependence $d(T)$; it can be seen that after the initial abrupt growth
the magnitude rapidly subsides.  For the values of $T>0.01$ no
significant oscillations can be observed: mean fields $\langle
x\rangle$ and $\langle y\rangle$ display only weak fluctuations around
constant values.  This indicates that the system has arrived at the
steady distribution; however, in contrast to the sharp steady
distribution at $T=0$, in the latter case the width of the
distribution is quite large. Of course, the stationary distribution
does not imply stationary states for individual units: since the noise
intensity is high, each of them remains in the regime of frequent
spiking.  However, coherence between the individual spike events is
lost, and the ensemble average displays no time dependency.

An increase of the coupling strength $\gamma$ produces a quantitative
change in this picture (Fig.\ref{dir_diam}b): now the spiking state
appears somewhat later and persists in a larger interval of the values
of $T$. Qualitatively we observe the similar resonance-like picture
with change of coupling strength: there exists a certain range of
noise intensities which allows for the spiking regime of the mean
field; both the too weak noise and the too strong one suppress the
spiking. However, the enlargement of the noise range where spiking
occurs proves that the array enhances the collective spiking.

The further growth of $\gamma$ leads to disappearance of the spiking
regime as well as of the state of minor oscillations: for $\gamma>2.2$ we
failed to observe significant temporal variations of the mean field
at arbitrary noise intensities. Apparently, too strong coupling 
among the elements counteracts the noise: it holds individual 
systems together in the vicinity of the equilibrium and
does not allow them to escape across the firing 
threshold\cite{foot_2}.

\begin{figure}[h]
\centerline{\epsfxsize=0.4\textwidth\epsfbox{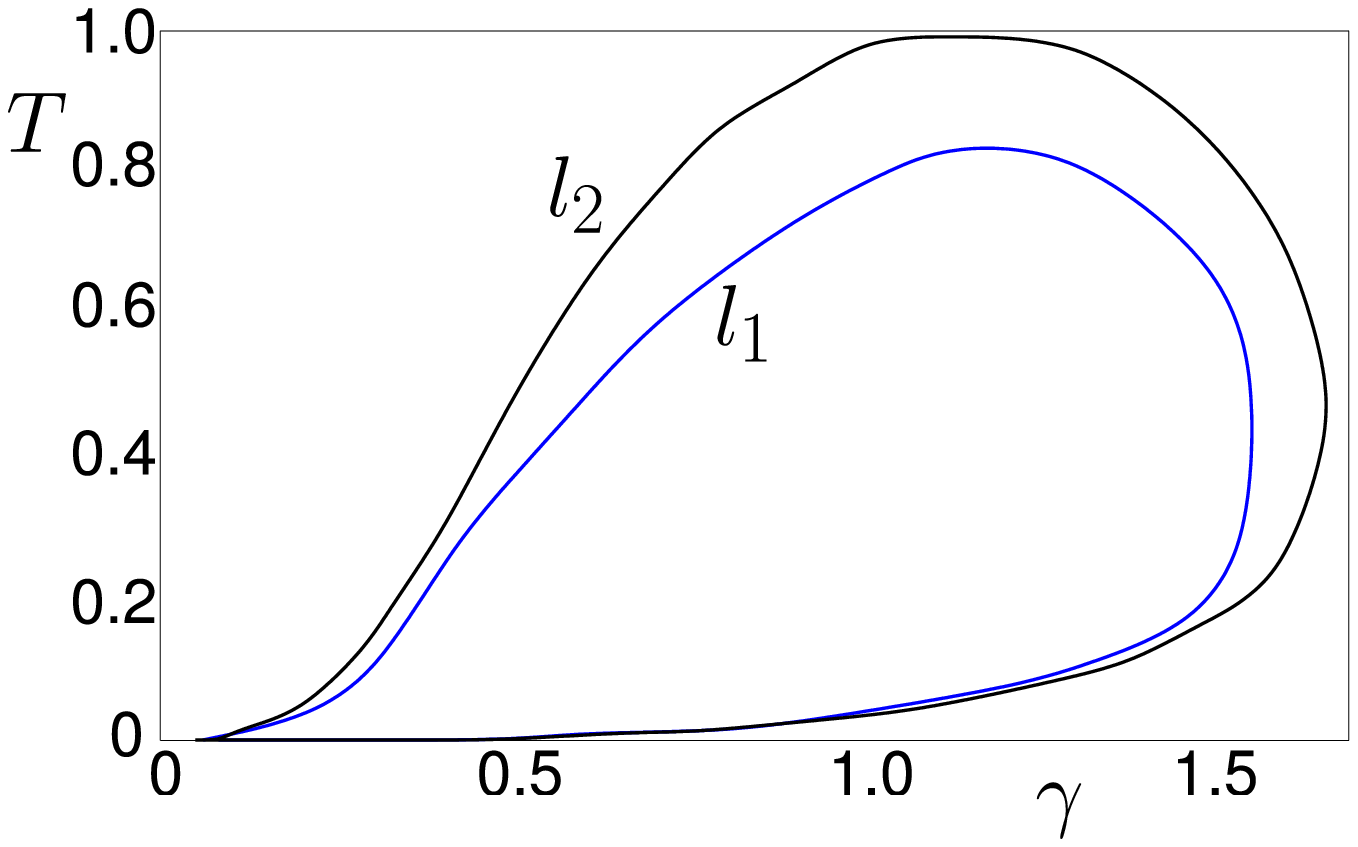}\quad
\epsfxsize=0.4\textwidth\epsfbox{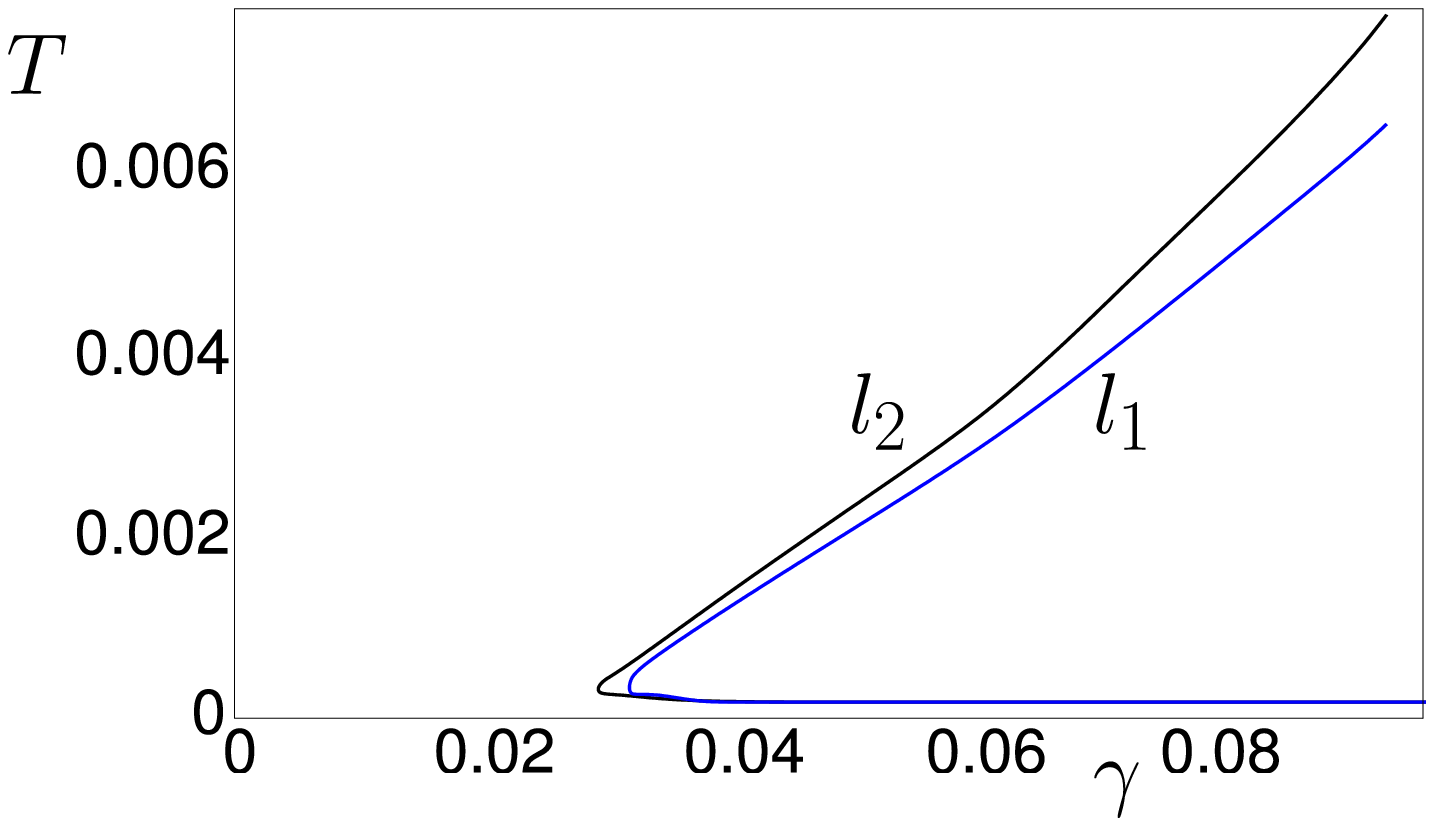}}
\caption{Domains of existence of nonstationary regimes on the 
coupling-strength $\gamma$ vs noise intensity $T$ plane: 
there are no spiking states outside the inner curve $l_1$ 
and no oscillatory states at all outside the outer curve
$l_2$. Left panel: global view; right panel: enlarged part for small 
values of $\gamma$, other parameters like in Fig. \ref{triport}.}
\label{dir_border}
\end{figure}

Boundaries of existence of nonstationary states on the parameter plane
are shown in  Fig.\ref{dir_border}. Here we see that not only the strong
but also the weak coupling does not benefit the spiking state:
apparently, for $\gamma <0.03$ the coupling is insufficient in
order to synchronize the firing events of individual elements.
It should be noted that the upper part of the curve $l_1$ is somewhat
ambiguous: here the spiking state arises from subthreshold oscillations
in the course of decrease of $T$ not via the abrupt transition, 
but rather through the continuous (although quite fast) growth of their 
amplitude; the plotted curve corresponds to the parameter values
which ensure that the magnitude $d$ equals 1.

Summarizing the results of numerical simulations, we state
that the nonstationary mean fields can be observed only in a restricted
range of the noise intensity, provided that the coupling strength does
not exceed a certain value. Among those nonstationary states we can
distinguish the chaotic subthreshold oscillations, the intermittent
spiking regime and the regular spiking pattern.

\section{Gaussian approximation: dynamics of cumulants}
\label{chapCumulants}
\subsection{Governing equations and geometry of phase space}

The more detailed analysis of the set of coupled systems requires
the deeper knowledge of the instantaneous distributions of $x_i$
and $y_i$. In the limit of $N\to \infty$ such knowledge can be
obtained from the analysis of the corresponding Fokker-Planck equation;
presence of the coupling term in Eq.(\ref{fhn}) adds to this
equation the quadratic nonlinearity and destroys its variational
character. Recently, the non-stationary Fokker-Planck equation for this 
problem was treated with the help of the expansion of the distribution density 
into the Hermite polynomials~\cite{acebron04}; several interesting
results were reported, but the global diagram of states is still missing. 

Description in terms of the moments of distribution turns the problem
into an infinite set of ordinary differential equations. In order
to keep the problem tractable, one should either truncate 
the set of moments or come up with a plausible closure hypothesis. 
We choose the latter way and approximate the simultaneous state of the
system by the Gaussian distribution of $x_i$ and $y_i$ with time-dependent
parameters. Since for the Gaussian distribution (and {\em only} for it)
all cumulants of the high order vanish, this assumption allows us
to reduce the infinite-dimensional problem to the low-dimensional model.

\begin{figure}[h]
\centerline{\epsfxsize=0.48\textwidth\epsfbox{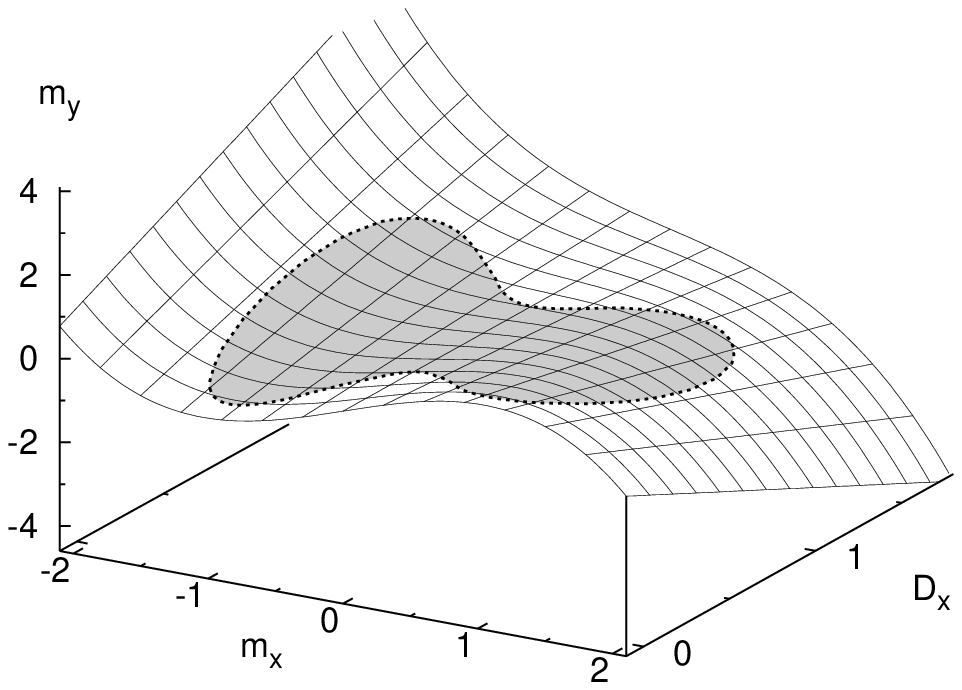}
\quad\epsfxsize=0.48\textwidth\epsfbox{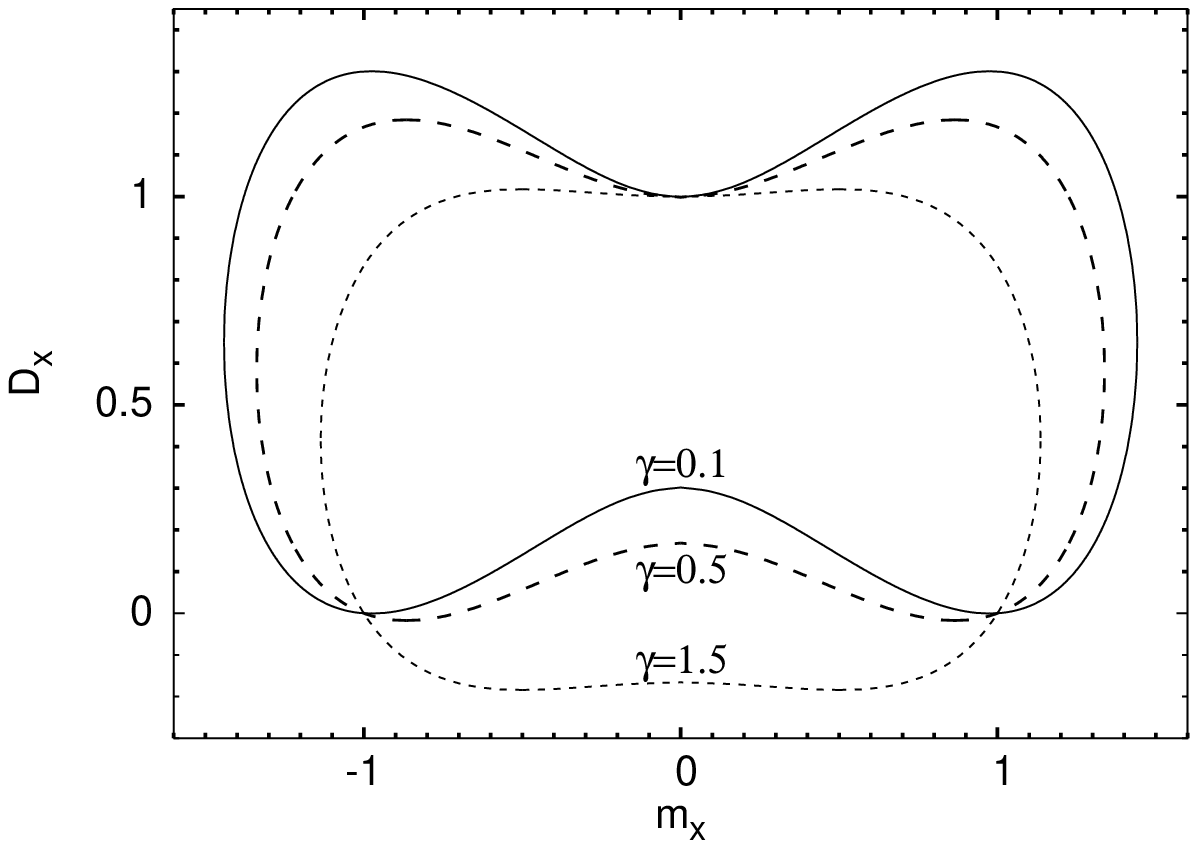}}
\caption{a) 3-dimensional Projection of the twodimensional slow surface
in the 5-dimensional phase space of the cumulant equations for $\gamma=0.1$. 
Shaded domain: repelling region where inequality (\ref{inequal}) holds.
b) Repelling region on the slow surface for $\gamma=0.1$, $\gamma=0.5$ 
and $\gamma=1.5$.}
\label{fig_slow}
\end{figure}

Let us average the evolution equations over the ensemble.
Conditions of vanishing of cumulants of the 3rd and 4th order provide
expressions for the higher-order moments $\langle x_i^3\rangle$,
$\langle x_i^2y_i\rangle$ etc. On substituting these expressions
into the averaged equations, we arrive at the dynamical system which
governs the behavior of the cumulants 
$m_x=\langle x_i\rangle$,  $m_y=\langle y_i\rangle$,
$D_x=\langle (x_i-m_x)^2\rangle$,  $m_y=\langle (y_i-m_y)^2\rangle$,
 $D_{xy}=\langle (x_i-m_x)(y_i-m_y)\rangle$
of the distribution:

\begin{eqnarray}
\label{cumulants}
\epsilon\frac{d}{dt}m_x & = & m_x - \frac{m_x^3}{3} - m_y  -m_x D_x \nonumber\\
\frac{d}{dt}m_y & = &  m_x+a  \nonumber\\
\epsilon\frac{d}{dt}D_x & = &  2D_x(1-D_x-m_x^2-\gamma)-2D_{xy} \\
\frac{d}{dt}D_y & = & 2(D_{xy}+T)     \nonumber\\
\epsilon\frac{d}{dt}D_{xy} & = &  D_{xy}(1-D_x-m_x^2-\gamma)-
D_y + \epsilon D_x         \nonumber
\end{eqnarray}

For small values of $\epsilon$ the separation of time scales in (\ref{cumulants})
turns $m_x$, $D_x$ and $D_{xy}$ into ``fast'' variables whereas $m_y$
and $D_y$ evolve on the slow time scale. Accordingly, in the 5-dimensional
phase space there is a two-dimensional surface which corresponds to slow
motions.

Location of this surface in the limit of vanishing
$\epsilon$ is obtained by setting $\epsilon=0$ in Eq.(\ref{cumulants}). 
It is convenient to parameterize the surface by coordinates $m_x$, $D_x$: 
given arbitrary values of these two variables 
(of course, only non-negative values of $D_x$ are physically meaningful) 
the three remaining coordinates are determined as
\begin{eqnarray}
m_y & = & m_x - m_x^3/3 - m_x D_x \nonumber\\
D_{xy} & = &  D_x(1-D_x-m_x^2-\gamma)\\
D_y & = &  D_x(1-D_x-m_x^2-\gamma)^2\nonumber
\label{slow_dyn}
\end{eqnarray}

Solving the linear system 
\begin{eqnarray}
 (1-D_x-m_x^2)\,\dot{m_x}& -\, m_x\,\dot{D_x} & =  m_x + a\nonumber\\
-(4m_x)\,D_x\,\dot{m_x} & +(1-3D_x-m_x^2-\gamma)\dot{D_x} & = 
  D_x\,+\,\frac{T}{1-D_x-m_x^2-\gamma}
\label{linsys}
\end{eqnarray}  
for variables $\dot{m_x}$ and $\dot{D_x}$,
yields evolution equations for dynamics {\em upon} the slow surface.
[Explicit expressions are straightforward, but too long to be quoted here].
Stability of this surface with respect to {\it transversal} perturbations 
is governed by the sign of the determinant of the system (\ref{linsys}):
slow surface is repelling for 
\begin{equation}
(m_x^2-1)^2+D_x(3D_x-4) < \gamma(1-m_x^2-D_x)
\label{inequal}
\end{equation}
and attracting otherwise; notably, position and shape of the repelling region
depend only on one of the parameters: the coupling strength $\gamma$.

One of the projections of the slow surface in the phase space is shown
in Fig.\ref{fig_slow}a; the butterfly-shaped repelling region is
shaded. The geometry of the phase space has implications for the
dynamics in general. We can expect that phase trajectories approach
the attracting part of the slow surface and move along it; in case of
penetrating into the region (\ref{inequal}), they move for a certain
time along the repelling surface (effect known as the ``canard
phenomenon''), depart from the slow surface and eventually approach
the attracting part again.

Indeed, the bifurcation analysis of Eq.(\ref{cumulants}) discloses
this picture.

\subsection{From stationary distribution to small-scale chaos}
We consider the equations (\ref{cumulants}) in the parameter range
$a>1$ where the solitary FitzHugh-Nagumo oscillator has a stable
steady state. For $T\geq 0$ the equations (\ref{cumulants}) also possess 
the unique steady solution in the domain $D_x,D_y\geq 0$:
\begin{eqnarray}
\label{steady}
m_x=-a,
&\;\;D_x=\frac{1-a^2-\gamma+\sqrt{(1-a^2-\gamma)^2+4T}}{2}& ,\nonumber\\
m_y=\frac{a^3}{3}-a+a\,D_x,&D_y=\epsilon D_x+T(a^2+D_x+\gamma-1),\\[1ex]
D_{xy}=-T\nonumber
\end{eqnarray}
(there exists also the
physically meaningless unstable state with negative $D_x$ and $D_y$).
This equilibrium describes the stationary distribution of oscillators; 
in the phase space the corresponding fixed point
is located on the slow surface. 

Destabilization of the equilibrium (\ref{steady}) under the increase
of the noise intensity has a convenient graphical interpretation. 
Recall that shape and position of the repelling region on the slow surface 
depend only on $\gamma$. Fig.\ref{fig_slow}b shows the outline of this 
region for several values of $\gamma$.  
As seen from Eq.(\ref{steady}),
the equilibrium value of $m_x$ is $T$-independent, whereas the value of $D_x$ 
is the monotonically growing function of $T$. 
At $T$=0 the fixed point lies on the abscissa of Fig.\ref{fig_slow}b.
For all values of $\gamma$, the border of the repelling region passes through 
the point $m_x$=--1, $D_y$=0; part of the abscissa to the left from this point 
always lies in the attracting region. 
Accordingly, for $a>1$ and arbitrary $\gamma$
the stationary solution is stable in the absence of noise
and (by continuity) at very small values of $T$.

\begin{figure}[h]
\centerline{\epsfxsize=0.5\textwidth\epsfbox{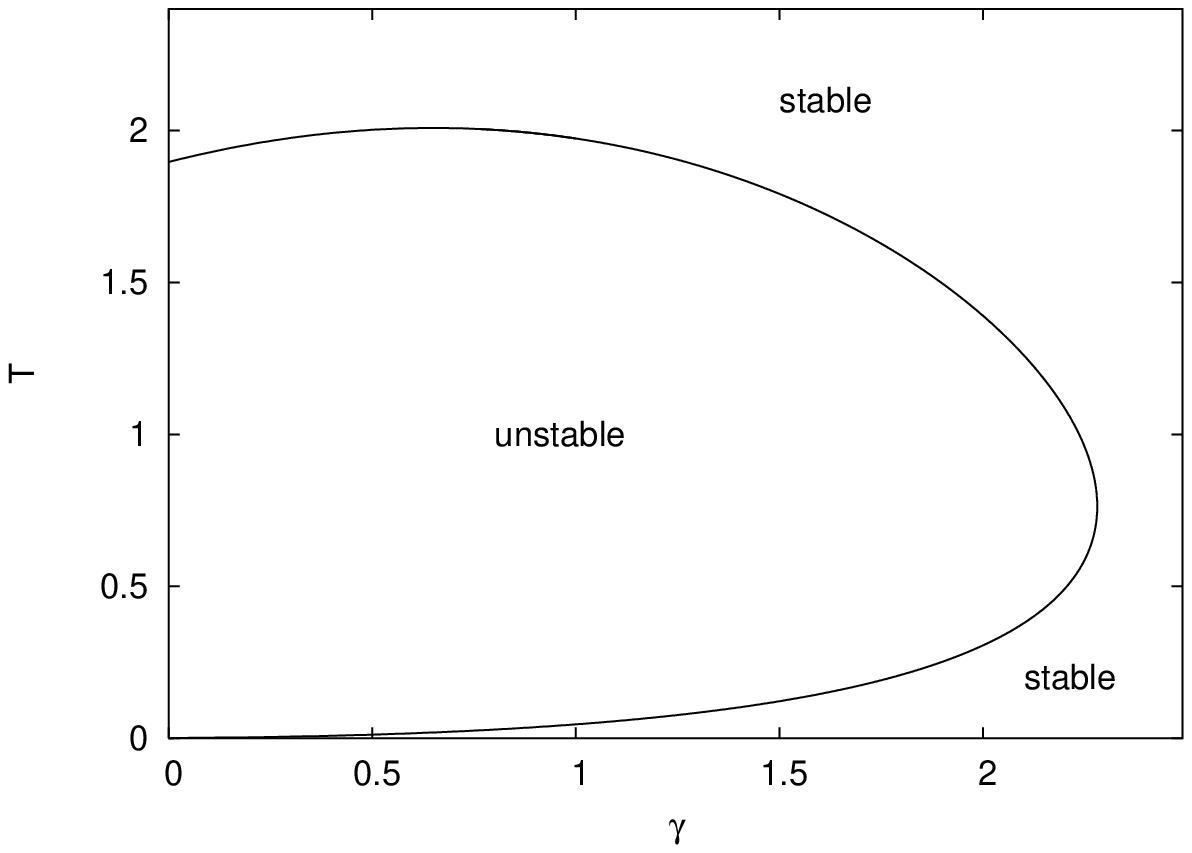}
            \epsfxsize=0.5\textwidth\epsfbox{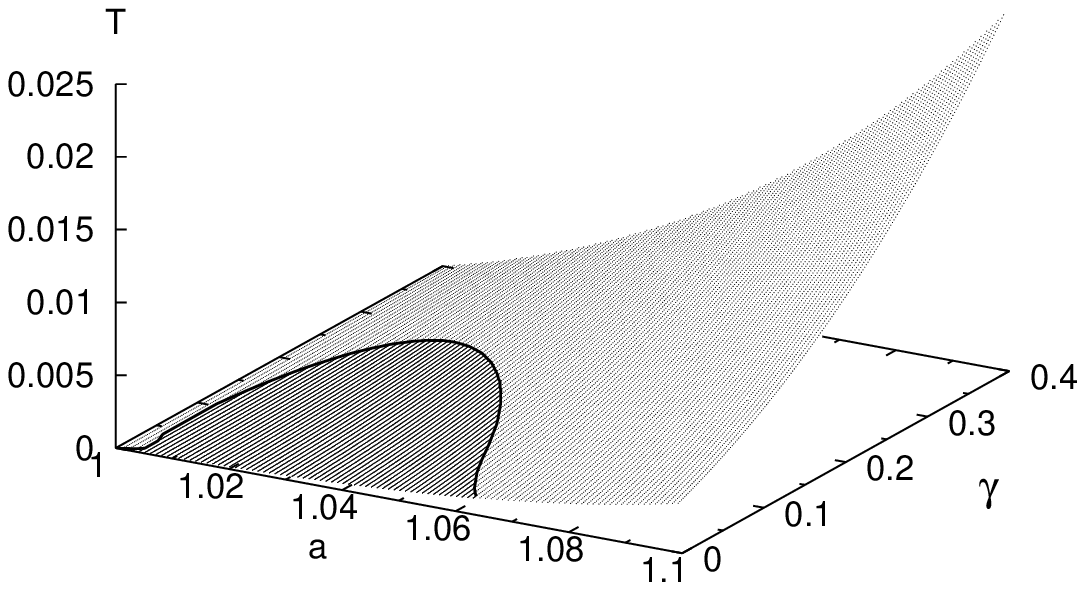}}
\caption{(a) State diagram of the equilibrium at $a=1.05$, $\epsilon=0.01$. 
Region of stability is bounded by the curve of the 
Andronov-Hopf bifurcation. (b) Bifurcation surface of the
Andronov-Hopf bifurcation in the parameter space of
Eqs(\ref{cumulants}) with $\epsilon=0.01$. The shaded region indicates
a subcritical bifurcation.}
\label{hopf}
\end{figure}

As a consequence of the increase of $T$, the fixed point 
moves upwards along the vertical line $m_x=-a$ 
and can cross the repelling region. Destabilization occurs 
when the fixed point hits the lower border of this region;
in terms of Eqs (\ref{cumulants}) this event is the Andronov-Hopf bifurcation. 
In the limit $\epsilon\to 0$ the critical value $T_H$ is given 
by the smaller root of the quadratic equation
\begin{equation}
9T^2+T(16b^2-16-12b-4\gamma+9b\gamma+2\gamma^2)+2b(b+\gamma)^2(2+2b+\gamma)=0
\label{quadratic}
\end{equation}
where $b\equiv a^2-1$.

To enable this instability, the vertical line $m_x=-a$ should cross
the repelling region in Fig.\ref{fig_slow}b.  In terms of the
parameter values this condition reads as $$\gamma\leq
\gamma_0=2(3a^2-1-2a\sqrt{3a^2-3}\,).$$ For $\gamma>\gamma_0$ the
fixed point on the slow surface is always located to the left of the
repelling region, and the equilibrium is stable for all values of $T$.
This can be interpreted as stabilization of the equilibrium
distribution by strong coupling. With the increase of $a$ the value of
$\gamma_0$ becomes smaller; it vanishes at
$a=a_0=\sqrt{1+\sqrt{4/3}}\approx 1.467$. For $a>a_0$ the steady state
is stable indendently of the values of $\gamma$ and $T$.  At small nonzero
$\epsilon$ the qualitative and, largely, the quantitative picture
persists; merely the values of $\gamma_0$ and $a_0$ slightly change.

\begin{figure}[h]
\centerline{\epsfxsize=0.9\textwidth\epsfbox{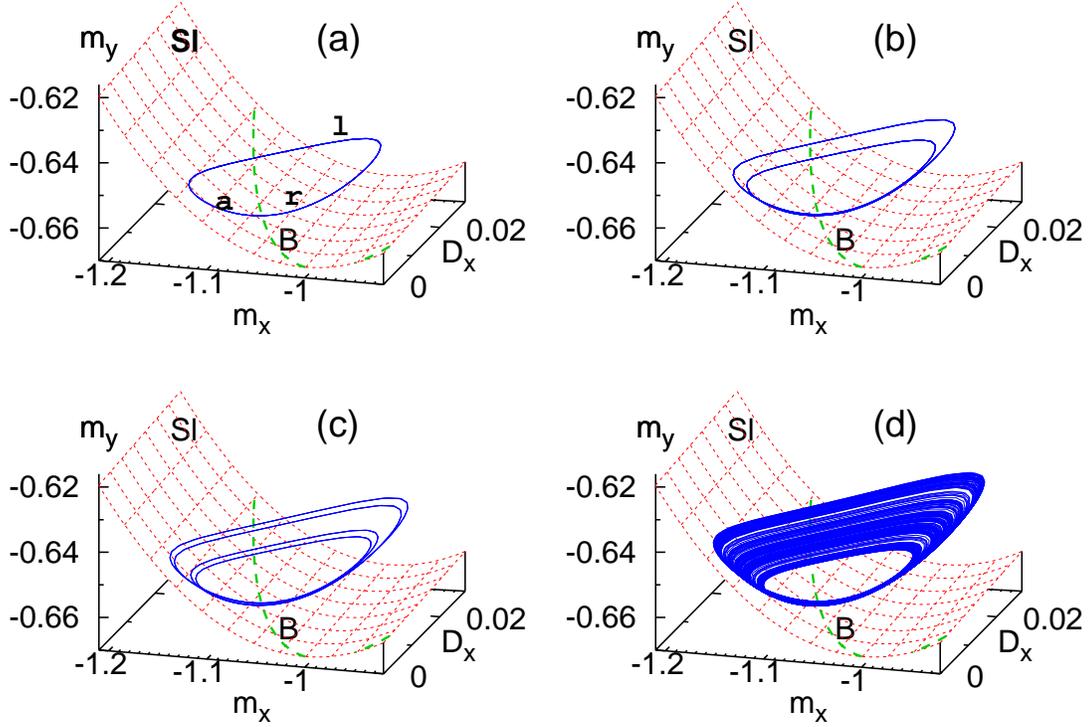}}
\caption{Projections of phase portraits for Eqs(\ref{cumulants}) 
(cf. Fig. \ref{fig_slow}) for different regimes of local
subthreshold oscillations. The dotted grid $Sl$ shows the slow
surface in the phase space. The surface $Sl$ is stable to the
left of the dashed line $B$. 
 (a) $T=0.00157$, (b)
$T=0.00158$, (c) $T=0.0015826$, (d) $T=0.001585$.  Other parameters:
$a=1.05$, $\gamma=0.1$, $\epsilon=0.01$}
\label{leap}
\end{figure}

Under sufficiently high values of $T$ the fixed point moves over the
upper border of the repelling region on the slow surface: the
equilibrium regains stability.  Again, this is the Andronov-Hopf
bifurcation; the asymptotics at $\epsilon\to 0$ for the corresponding
value of $T$ is given by the larger root of Eq.(\ref{quadratic}).
Shape of the bifurcation curve on the plane of parameters $\gamma$ and
$T$ is shown in Fig.\ref{hopf}a.

Thus we observe that in the domain of low and moderate values 
of $\gamma$ the equilibrium exhibits two Andronov-Hopf bifurcations:
the destabilizing and the stabilizing one.
Computation of the cubic term in the amplitude expansion 
discloses a relatively small area on the parameter plane of $a$ and $\gamma$
where the first bifurcation is subcritical (Fig.\ref{hopf}b):
the unstable periodic solution branches off in the direction of
small $T$. Here, a hysteresis occurs: the time-independent regime coexists with 
the stable oscillatory state; the latter is born from the tangent bifurcation 
shortly before the destabilization of the equilibrium. 
For most of the values of $a$ and $\gamma$,
however, the first  Andronov-Hopf bifurcation is supercritical: 
the stable limit cycle with small amplitude 
$\sim \sqrt{T-T_H}$ and frequency $\sim\epsilon^{-1/2}$ is born. 
The second bifurcation is always supercritical: the stable limit cycle
is born into the domain of lower values of $T$. With regards to 
finite-amplitude states, there seem to be no non-decaying solutions 
in the part of the parameter plane above the bifurcation curve in 
Fig.\ref{hopf}a. 

Near the lower branch of the bifurcation curve, the newborn limit
cycle includes the segment (denoted by $a$ in Fig.\ref{leap}a) which
lies close to the attracting region of the slow surface, and the
segment $r$ which leads along the repelling region; from there, the
orbit leaps back to the attracting region (segment $l$ in
Fig.\ref{leap}a).

Since the slow surface is two-dimensional, one can expect new
qualitative effects in comparison to the single FitzHugh-Nagumo
system with its $S$-shaped one-dimensional slow manifold.  Indeed,
when $T$ is further increased, the periodic orbit loses stability by
means of the period-doubling bifurcation.  This event is followed by
the sequence of further period-doublings which culminates in the onset
of the chaotic state. In terms of the ensemble, the latter describes
minor (localized) chaotic oscillations around the stationary
distribution. Transformation of the attracting orbit and its position near
the slow surface is depicted in Fig.\ref{leap};

\begin{figure}[h]
\centerline{\epsfxsize=0.6\textwidth\epsfbox{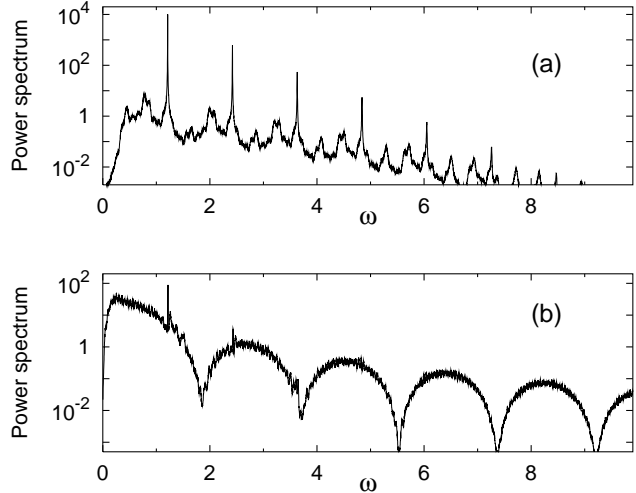}}
\caption{Transition in the power spectrum of the cumulant equations.
$a=1.05$, $\gamma=0.1$, $\epsilon=0.01$. (a) $T$=0.001585, (b)
$T$=0.001586.  }
\label{spectra_cum}
\end{figure}

The power spectrum of a trajectory on the chaotic attractor consists
of the peak at the characteristic frequency of the minor oscillations,
its harmonics and the broad noisy background (Fig.\ref{spectra_cum}a).

\subsection{From chaotic to regular spiking}
Further growth of $T$ brings about dramatic changes in the
shape and size of the attractor. As mentioned above, attracting orbits
involve segments which lie close to the repelling part of the slow surface.
In such situation a minute variation of the parameter may be a reason
for the abrupt (albeit continuous) increase of the attractor size 
by several orders of magnitude: the so-called ``canard explosion''.
This phenomenon has been well documented for the case when the attractor
is a limit cycle (e.g. in the forced Van der Pol equation~\cite{Glendinning}, 
and in the context of a single FitzHugh system disturbed
by noise~\cite{makarov01,volkov03,marino04}).
The peculiarity of our situation is that
the ''canard explosion'' happens not to an individual periodic orbit
but to the chaotic attractor as a whole (of course, individual unstable
periodic orbits embedded into the attractor also experience the explosion,
but this remains unnoticed for an observer who watches only the
stable sets). 

\begin{figure}[h]
\centerline{\epsfxsize=0.8\textwidth\epsfbox{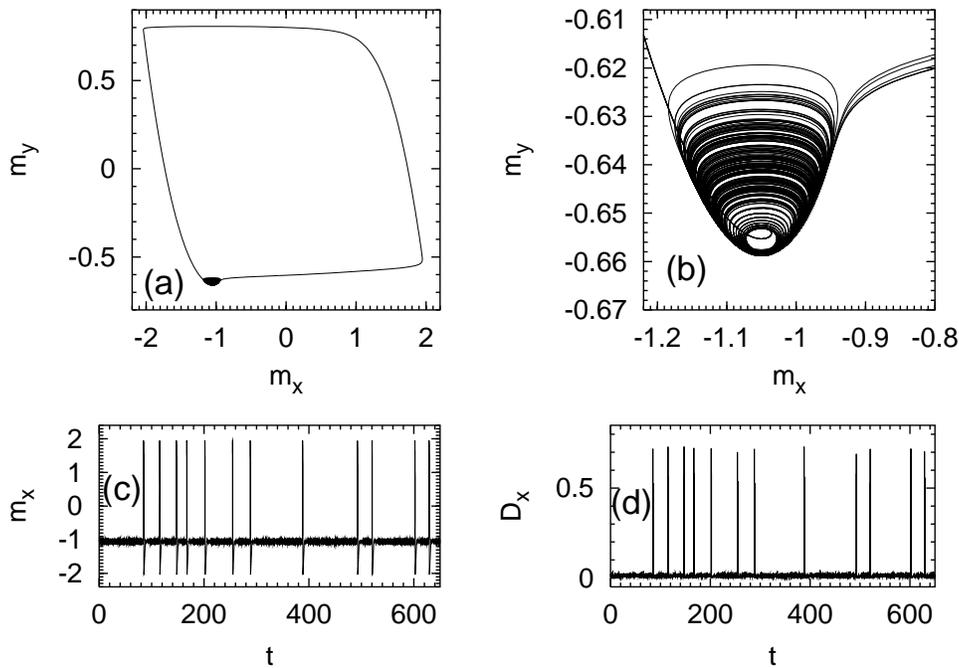}}
\caption
{Regime of irregular spiking in Eqs(\ref{cumulants}); $a$=1.05,
$\gamma$=0.1, $\epsilon$=0.01, $T$=0.001586. (a) projection of phase
portrait; (b) enlarged part of phase portrait; (c) plot of $m_x(t)$;
(d) plot of $D_x(t)$. }
\label{spiking_a}
\end{figure}

Projections of phase portraits and temporal evolution of the mean fields 
$m_x$ and $m_y$ in this state 
are shown in Fig.\ref{spiking_a}. Apparently, this regime corresponds to
intermittent chaotic spiking: after many minor oscillations in the
vicinity of the unstable steady equilibrium, the system exhibits
a spike of large amplitude, followed by the next epoch of localized
chaotic oscillations.

\begin{figure}[h]
\centerline{\epsfxsize=0.5\textwidth\epsfbox{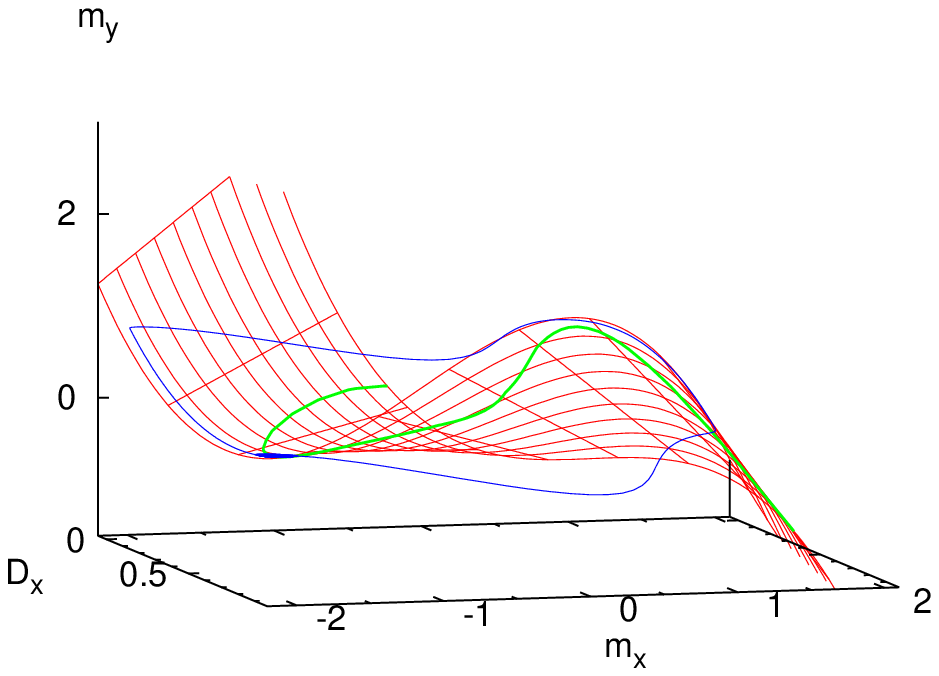}
            \epsfxsize=0.5\textwidth\epsfbox{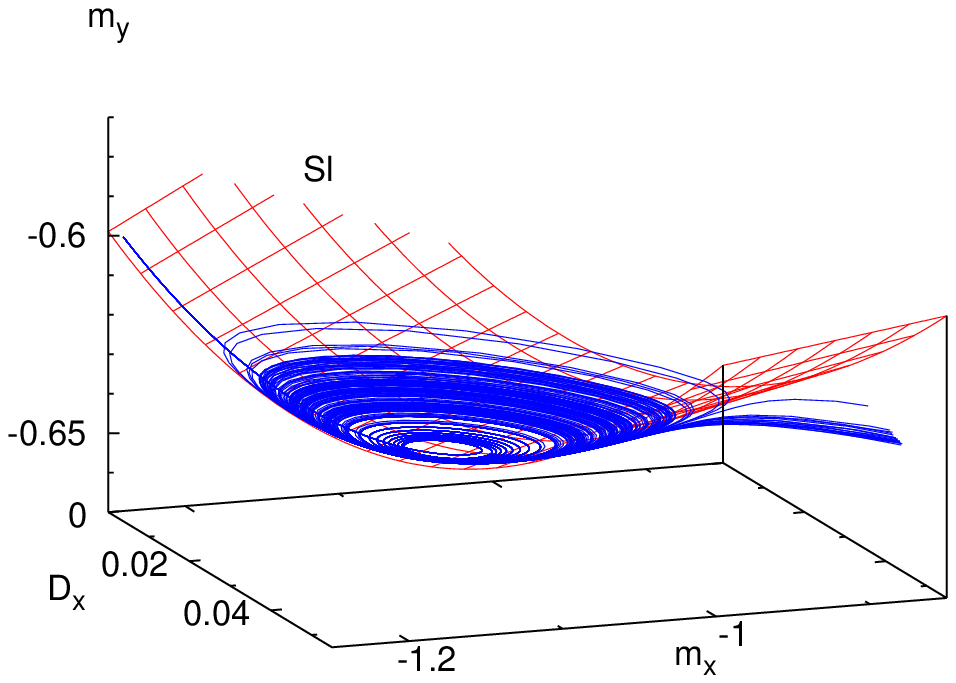}}
\caption{Projection of the attractor and the slow surface 
    in the state of irregular spiking.  Left: global view of the attractor;
        right: enlargement of the domain near the unstable
          equilibrium.}
\label{can_3d}
\end{figure}

\begin{figure}[b]
\centerline{\epsfxsize=0.6\textwidth\epsfbox{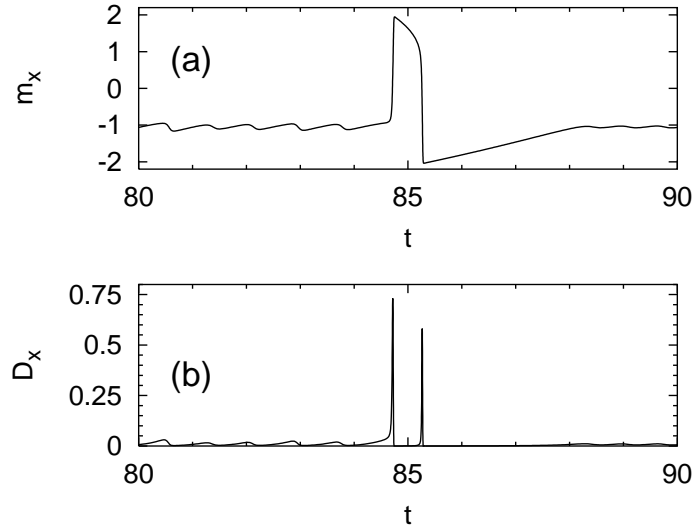}}
\caption{Time evolution of the mean field and its variance within a single 
spike.   (a) variable $m_x(t)$; (b)
variable $D_x(t)$. Parameter values like in Fig.\ref{spiking_a}.
Rapid changes of $m_x$ in either direction induce 
peaks of $D_x$.}
\label{spiking_b}
\end{figure}

A joint graphical presentation of the attractor and the slow surface
is helpful for understanding the nature of spiking. The enlargement of
the domain around the equilibrium in Fig.\ref{can_3d}a discloses that
each arrival of a trajectory to this domain occurs via a steep
descent into the ``valley'' along the slow surface. This return is
followed by several minor revolutions, which share the familiar
pattern: two segments along the, respectively, attracting and the
repelling regions of the slow surface, and a return back to the
attracting region. Finally, the trajectory ``gets through'' the slow
surface \cite{foot_3} and is repelled far away from the equilibrium. A
global picture (Fig.\ref{can_3d}b) indicates that this flight ends in
the domain of positive $m_x$ where the trajectory rapidly approaches
the slow surface and moves along it for a certain time; the spike is
accomplished by another fast flight back to the initial domain.

Accordingly, each single spike of the mean field 
is accompanied by two spikes of the variance (Fig. \ref{spiking_b}):
the latter remains small on each of two slow segments
but rapidly peaks during each of two fast flights.

\begin{figure}[h]
\centerline{\epsfxsize=0.55\textwidth\epsfbox{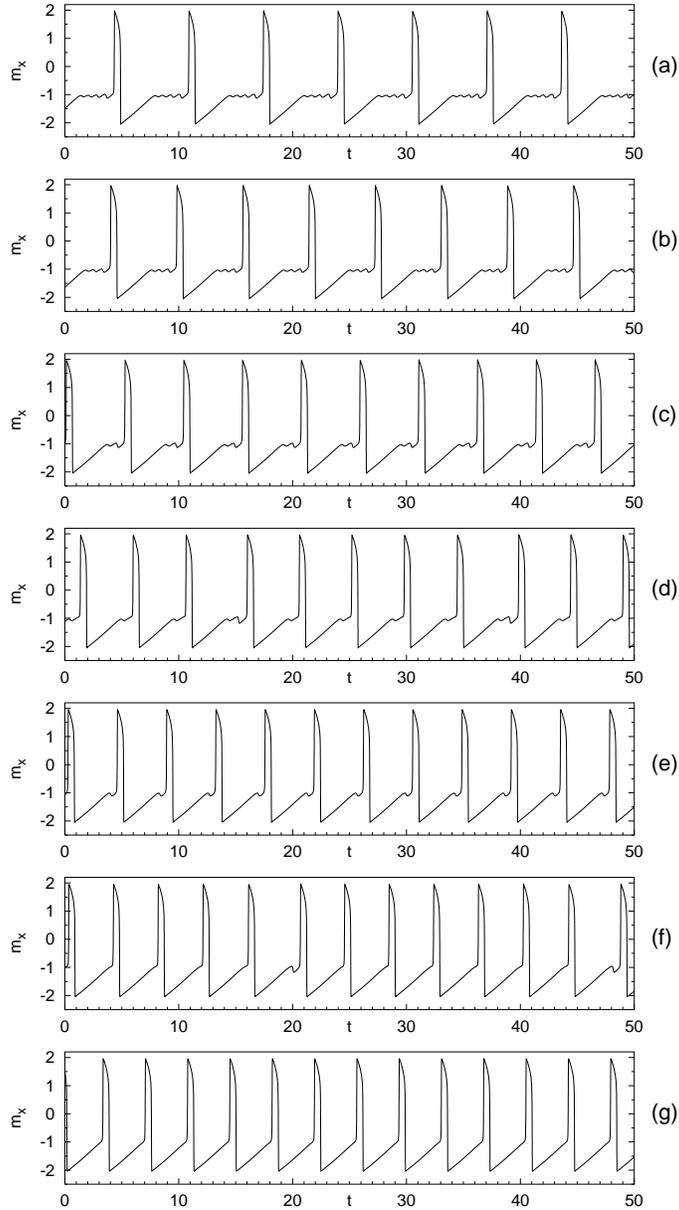}}
\caption{Transition from intermittent to regular spiking in 
the cumulant equations with $\epsilon=0.01$, $a=1.05$, $\gamma=0.1$:
(a) $T=0.00168$, (b) $T=0.00172$, (c) $T=0.00185$, (d) $T=0.0018569$,
(e) $T=0.00220$, (f) $T=0.0023105$, (g) $T=0.0024$. }
\label{reg_spike_cum}
\end{figure}

Transition to the spiking regime is accompanied by sharp changes in
the power spectrum of the process; this is shown in
Fig.\ref{spectra_cum}b.

\begin{figure}[h]
\centerline{\epsfxsize=0.49\textwidth\epsfbox{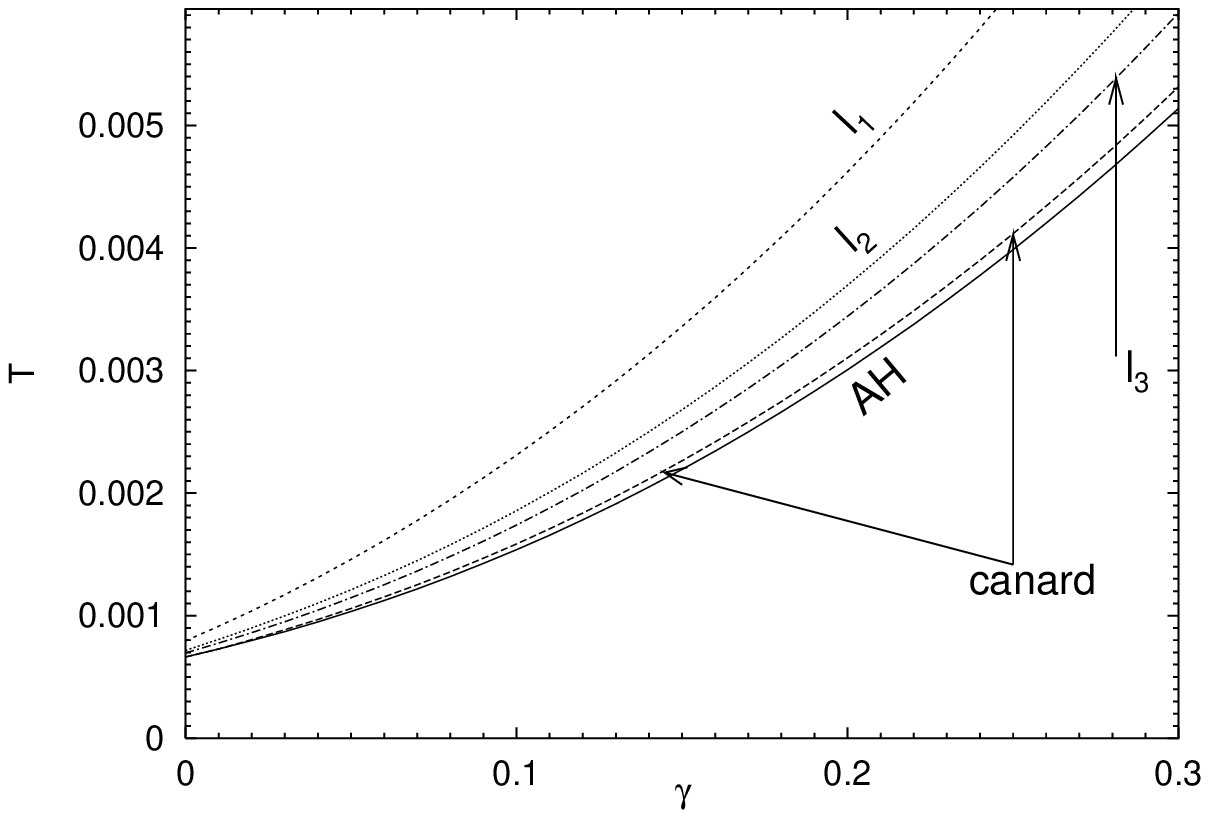}\quad
\epsfxsize=0.49\textwidth\epsfbox{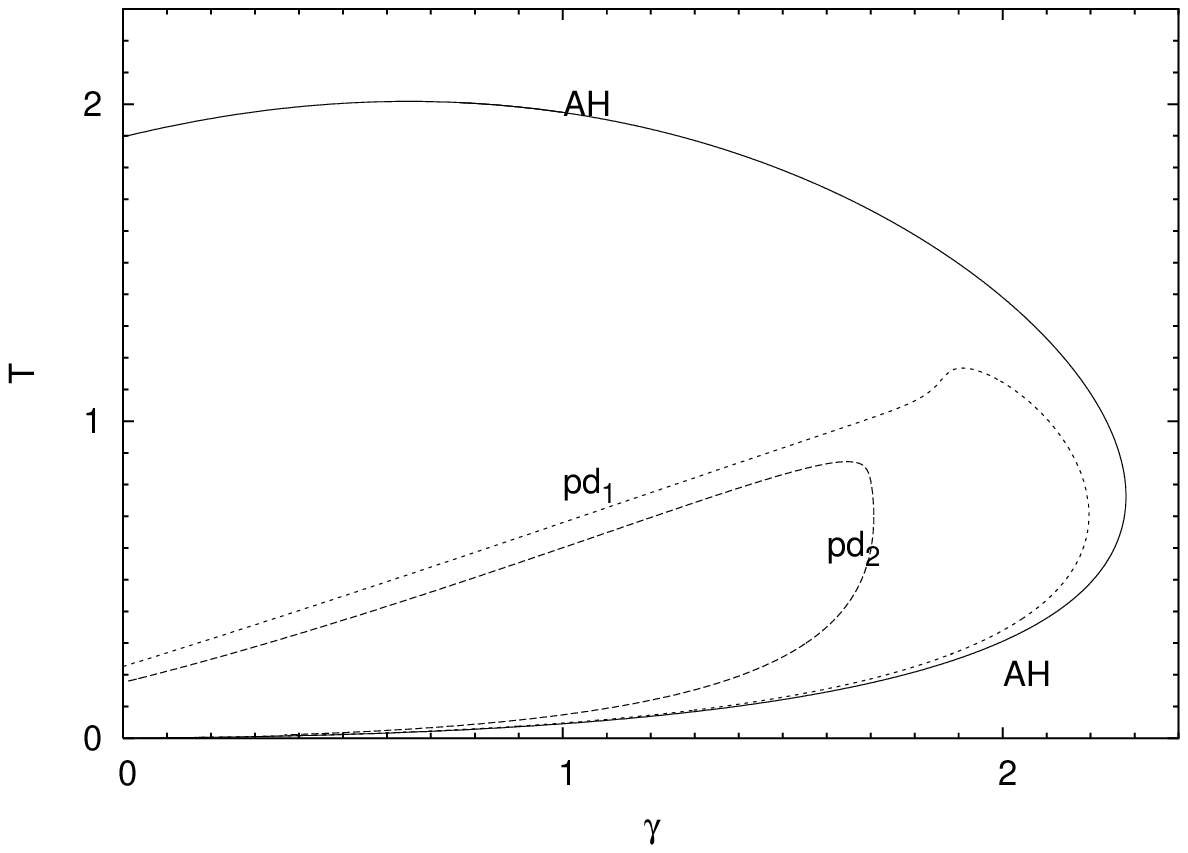}}
\caption
{a) Part of the bifurcation diagram for cumulant equations at $a=1.05$,
$\epsilon=0.01$ for small values of the coupling $\gamma$.
$AH$: Andronov-Hopf bifurcation; Canard explosion: onset of intermittent
spiking oscillations, $l_3$: onset of regime with 2 minor oscillations
between spikes, $l_2$: onset of regime with 1 minor oscillation
between spikes, $l_1$: onset of regime without minor oscillation
between spikes\\
b) Important transitions on the parameter plane: $AH$:
Andronov-Hopf bifurcation; $pd_{1,2}$: period-doubling bifurcations.}
\label{bd_cum}
\end{figure}

States with chaotic spiking occupy only a narrow strip in the
parameter space. Further increase of $T$ regularizes the spiking
pattern: individual spikes are now separated by equal amounts of minor
oscillations. With growth of $T$ this amount gets smaller: it is a kind
of the inverse ``period-adding'' sequence in which $N$ localized
oscillations between two subsequent spikes are replaced by $N$--1 ones
(there is also a narrow transition region with more complicated
states), then $N$--1 yields to $N$--2, etc \cite{Kaneko83}. Finally the minor
oscillations completely disappear and the system proceeds into the
regime of uninterrupted periodic spiking. Different stages of this
process are shown in Fig.\ref{reg_spike_cum} whereas the basic
transition curves on the bifurcation diagram are plotted in
Fig.\ref{bd_cum}a.

\subsection{Strong noise: back from regular spiking to stationary state}
As discussed above, under high values of $T$ the only attractor of the
system is the stationary state. Formally, it is the same solution
(\ref{steady}) as in the case of low $T$; however, now this state
describes the broad stationary distribution with high values of $D_x$
and $D_y$. Here we briefly outline the way to this stationary state
from the regular spiking pattern.

\begin{figure}[h]
\centerline{\epsfxsize=0.7\textwidth\epsfbox{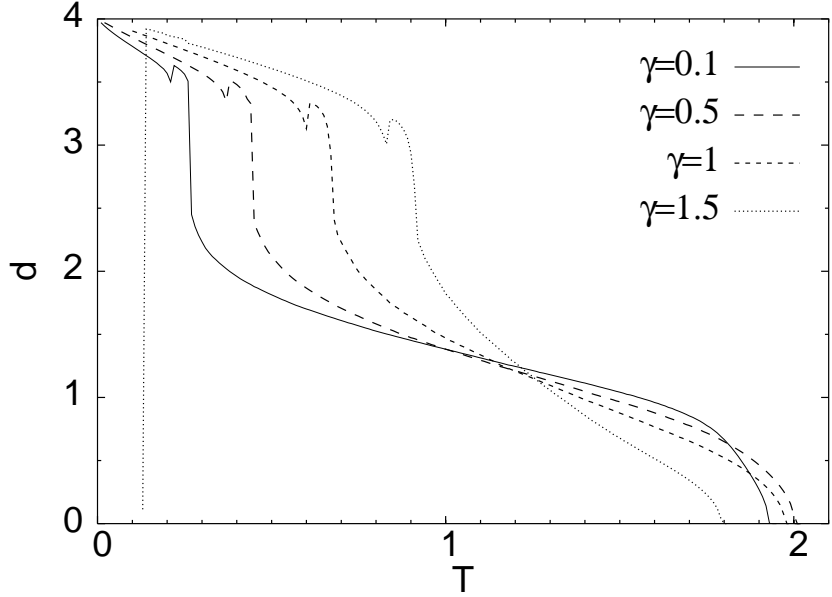}}
\caption{Resonance like dependence of the magnitude of oscillations as function of the  noise intensity $T$. Results of integration of the cumulant equations.}
\label{diameter}
\end{figure}

Quantitatively, we characterize the oscillations in terms of the same
``magnitude'' as in Sect.\ref{chapLangevin}: now this ``attractor
diameter'' is defined as $d\equiv \max(m_x)-\min(m_x)$.  Dependence of
$d$ on $T$ for several fixed values of $\gamma$ is shown in
Fig.\ref{diameter}.

We observe that the weak decay of the amplitude at low values of $T$
is interrupted by the sharp decrease by a factor of $\sim 2$ which
occurs within the narrow interval of $T$. This decrease is followed by
the final interval of continuous decay until the periodic state
disappears in the Andronov-Hopf bifurcation; here, the segments of the
curves obey the square-root law. The sharp decrease reminds of the
canard explosion at low values of $T$, although it is less
dramatic. At a closer look we see that the decrease is always preceded
by a short interval in which $d(T)$ is non-monotonic. Checking the
corresponding parameter values, we disclose here additional
bifurcations.  The local minimum of $d(T)$ corresponds to the
period-doubling which occurs on the upper branch of the curve $pd_2$
on the parameter plane in Fig.\ref{bd_cum}b.

This event is followed by the whole period-doubling scenario and the
narrow range of $T$ in which the spiking pattern turns chaotic
again. In contrast to the previously described intermittent chaotic
spiking, there are no small oscillations between the subsequent
spikes; only the amplitudes of the spikes differ.  The magnitude $d$
weakly grows again, but in the course of the further minor increase of
$T$ the loops of the orbit ``slide off'' the broad part of the
repelling region of the slow surface, and $d$ rapidly decreases.
Meanwhile, the chaotic state turns back into the regular one, and
finally the reverse period-doubling bifurcation restores the simple
periodic state. On the parameter plane the curve of the latter
bifurcation turns out to be the upper branch of the already familiar
period-doubling curve described in the previous subsection (line
$pd_1$ in Fig. \ref{bd_cum}b).  Fast ``halving'' of the attractor
diameter takes place just below the curve $pd_1$; in the broad parameter
range above this curve and below the upper branch of the curve $AH$ of
the Andronov-Hopf bifurcation the amplitude of the limit cycle slowly
decays.

For the values of $\gamma$ to the left of the curve $pd_2$
$(1.7<\gamma<2.2)$ the ordering of states is slightly different. Here
the reverse period-adding sequence which simplifies the intermittent
spiking pattern at low values of $T$, ends up in a limit cycle with
several (4-6) spikes which exists over the broad range of
$T$. Slightly below the upper branch of the curve $pd_1$ 
the transition to chaos occurs;
it is followed by the reverse sequence of period-doubling bifurcations
and very strong decrease of the diameter $d$.

Summarizing the results obtained from the analysis of the equations
which govern the dynamics of the cumulants, we observe their
remarkable qualitative correspondence with the basic knowledge
obtained from the integration of the Langevin equations. Both
the very weak and the very strong noise correspond to stationary
probability distributions. Description in terms of the cumulants also
confirms the stabilization of the stationary distribution by the
strong coupling.  The attractors of the cumulant equations practically
do not differ from the phase portraits of the mean fields computed
from simulations of the Langevin equations; the patterns of power spectra, 
especially for
the state of intermittent spiking, also look strikingly similar.  Besides,
the analysis of the cumulant equations discloses the detailed picture
of transitions between different non-stationary states: birth of the
intermittent chaotic spiking from chaotic subthreshold oscillations
and the subsequent regularization of the spiking pattern under the
action of noise.  Quantitatively, in the cumulant equations the
critical values of the noise intensity which correspond to transitions
between different states, turn out to be noticeably higher than those
in the numerical simulations of large ensembles of
FitzHugh-Nagumo systems.

\section{Discussion and Conclusions}
\label{chapConclusions}

We have investigated an ensemble of globally coupled FitzHugh-Nagumo
elements under the action of additive Gaussian white noise. 
With the help of numerical simulations of stochastic equations
we have verified that noise intensity becomes 
an additional control parameter of the system. By changing noise intensity
we  provoke qualitative transformations of the global response of the system, 
characterized by the mean field.

In order to gain better understanding of dynamical mechanisms of these 
transitions we employed an approximate description in terms of cumulants 
of the Gaussian distribution. 
On the whole, modeling a large ensemble of globally coupled excitable
units by the Gaussian distribution with time-dependent characteristics
proved to be a fruitful approach: it has reduced the problem to the
low-dimensional deterministic dynamical system which, in the large
part of the parameter space, reproduced basic properties of the
underlying stochastic ensemble.  This allows us to conjecture that the
fine details of the transitions between the states, accessible only in
the framework of this model, keep relevance also for the whole
ensemble.

In our example only the slow variables have been subjected to the additive
noise. Of course, in the natural environment and in the laboratory
experiments one cannot reduce the influence of fluctuations to
only the slow variables: the fast ones can be affected as well. This can be
accounted for, if the additive sources of Gaussian noise are included
into the first of Eq.(\ref{fhn}). We performed a set of numerical
experiments with such equations, and observed, in general, the states
similar to those described in the previous sections: steady
distributions at high or very low noise intensities, subthreshold
oscillations and spiking states in the intermediate range. The
analysis of the respective cumulant equations can be pursued along the
same lines as above.  Since the noise acts on the fast variables, the 
shape and position of the repelling part of the slow surface depend 
now on the noise intensity; qualitatively, however, the only new effect in
comparison with those described in Sect.\ref{chapCumulants} seems to
be the direct transition from periodic subthreshold oscillations to
regular spiking without the mediation of chaotic states; this is, of
course, the usual canard explosion.

Of course, the Gaussian distribution is merely an approximation, and
its applicability has certain restrictions.  Thus, we failed to obtain
the quantitative correspondence for small values of the coupling strength
$\gamma$. Stochastic simulations indicate the absence of nonstationary
regimes in this parameter range (cf. Fig.\ref{dir_border}b) whereas
the analysis of the cumulant equations predicts the whole bifurcation
scenario for arbitrarily low values of $\gamma$ and even at $\gamma=0$
(cf. Fig.\ref{bd_cum}a). 
The higher the value of $\gamma$, the better the correspondence
between the behavior of the stochastic equation (\ref{fhn}) and the
deterministic model (\ref{cumulants}). However, since the
considered coupling is linear, an increase of its strength
reduces the complexity of dynamics
towards  a linear system. Hence, under the strong coupling  
the Gaussian approximation becomes a more adequate model of the
process, but the process itself gets simpler: there is little place
for nonlinear events and, hence, variation of $T$ fails to invoke
transitions in the behavior of the mean field.

{\bf Acknowledgement} This work was supported by SFB-555, 
an international cooperation grant from the DAAD 
(D/0104610) and the NSF (INT-0128974).


\begin{thebibliography}{99}
\bibitem{zaks05} We also performed simulations with noise terms 
in the activator equations, and obtained qualitatively similar
results (see the concluding section).
\bibitem{foot_2}
For the unbounded noise such rare large 
events should occur within sufficiently long time intervals, but
such intervals by far exceed our numerical capabilities.
\bibitem{foot_3}
Of course, this is only an optical illusion, a projection effect: 
a two-dimensional surface does not divide the 5-dimensional space.
\end{thebibliography}
\end{document}